\documentclass[11pt,leqno]{article}

\usepackage{amsmath}
\usepackage{amsthm}
\usepackage{amssymb}
\usepackage[dvips]{graphicx,color}

\theoremstyle{plain}
\newtheorem{theorem}{Theorem}[section]
\newtheorem{lemma}[theorem]{Lemma}
\newtheorem{proposition}[theorem]{Proposition}

\theoremstyle{definition}

\theoremstyle{remark}
\newtheorem{remark}[theorem]{Remark}

\numberwithin{equation}{section}

\begin{document}

\title{\textbf{Lipschitz continuity and monotone decreasingness of the solution to the BCS gap equation for superconductivity}}

\author{Shuji Watanabe\\
Division of Mathematical Sciences\\
Graduate School of Engineering, Gunma University\\
4-2 Aramaki-machi, Maebashi 371-8510, Japan\\
Email: shuwatanabe@gunma-u.ac.jp
\and Ken Kuriyama\\
Faculty of Education, Bukkyo University\\
Kyoto 603-8301, Japan\\
Email: kuriyama@bukkyo-u.ac.jp\\}

\date{}

\maketitle

\begin{abstract}
In the preceding paper \cite{watanabe3}, it is shown that the solution to the BCS gap equation for superconductivity is continuous with respect to both the temperature and the energy under the restriction that the temperature is very small.  Without this restriction, we show in this paper that the solution is continuous with respect to both the temperature and the energy, and that the solution is Lipschitz continuous and monotonically decreasing with respect to the temperature.

\medskip


\medskip

\end{abstract}

\section{Introduction and main result}
In this paper we study the temperature dependence of the nonzero solution to the BCS gap equation \cite{bcs, bogoliubov} for superconductivity:
\begin{equation}\label{eqn:bcsgapeq}
u(T,\,x)=\int_0^{\hslash\omega_D}
\frac{U(x,\,\xi)\, u(T,\, \xi)}{\,\sqrt{\,\xi^2+u(T,\, \xi)^2\,}\,}\,
\tanh \frac{\,\sqrt{\,\xi^2+u(T,\, \xi)^2\,}\,}{2T}\, d\xi,
\end{equation}
where the solution $u$ is a function of the absolute temperature $T \geq 0$ and the energy $x$
$(0 \leq x \leq \hslash\omega_D)$, and the constant $\omega_D>0$ (resp. $\hslash$) stands for the Debye angular frequency (resp. Planck's constant divided by $2\pi$). The potential $U$ satisfies $U(x,\,\xi)>0$ at all $(x,\,\xi) \in [0, \, \hslash\omega_D]^2$. In this connection, see Kuzemsky \cite{kuzemsky} for an interdisciplinary review.

The integral with respect to the energy $\xi$ in \eqref{eqn:bcsgapeq} is sometimes replaced by the integral with respect to the wave vector of an electron over the three dimensional Euclidean space $\mathbb{R}^3$. The existence and uniqueness of the solution to the BCS gap equation were established for $T=0$ by Odeh \cite{odeh}, and Billard and Fano \cite{billardfano}, and for $T\geq 0$ by Vansevenant \cite{vansevesant}. Bach, Lieb and Solovej \cite{bls} studied the gap equation in the Hubbard model for a constant potential and showed that the solution to this equation is strictly decreasing  with respect to the temperature. Recently, Frank, Hainzl, Naboko and Seiringer \cite{fhns} studied the asymptotic behavior of the transition temperature (the critical temperature) at weak coupling. Hainzl, Hamza, Seiringer and Solovej \cite{hhss} proved that the existence of a positive solution to the BCS gap equation is equivalent to the existence of a negative eigenvalue of a certain linear operator, and showed the existence of a transition temperature. Hainzl and Seiringer \cite{haizlseiringer} obtained upper and lower bounds on the transition temperature and the energy gap for the BCS gap equation.

The existence and uniqueness of the solution were established for each fixed temperature $T$ in the previous literature, and so the temperature dependence of the solution is not covered except for the work by Bach, Lieb and Solovej \cite{bls}. As is well known, studying the temperature dependence of the solution to the BCS gap equation is very important in condensed matter physics. This is because studying the temperature dependence of the solution by dealing with the thermodynamical potential leads to a mathematical proof of the statement that in the BCS model the transition to a superconducting state is a second-order phase transition. So it is highly desirable to discuss the temperature dependence of the nonzero solution to the BCS gap equation \eqref{eqn:bcsgapeq}.

To this end we define a nonlinear integral operator $A$ by
\begin{equation}\label{eqn:ouroperator}
Au(T,\,x)=\int_0^{\hslash\omega_D}
\frac{U(x,\,\xi)\, u(T,\, \xi)}{\,\sqrt{\,\xi^2+u(T,\, \xi)^2\,}\,}\,
\tanh \frac{\,\sqrt{\,\xi^2+u(T,\, \xi)^2\,}\,}{2T}\, d\xi.
\end{equation}
Here the right side of this equality is exactly the right side of the BCS gap equation \eqref{eqn:bcsgapeq}. Our nonlinear integral operator $A$ is defined on the sets $V_T$, $V$ and $W$ specified later. Since the solution to the BCS gap equation is a fixed point of the operator $A$ , we apply fixed point theorems to the operator $A$ and study the temperature dependence of  the nonzero solution to the BCS gap equation \eqref{eqn:bcsgapeq}.

Let $U_1>0$ is a  positive constant and set $\displaystyle{ U(x,\,\xi)=U_1 }$ at all $(x,\,\xi) \in [0, \, \hslash\omega_D]^2$. Then the solution to the BCS gap equation depends on the temperature $T$ only, and we denote the solution by $\Delta_1$. In this case the BCS gap equation \eqref{eqn:bcsgapeq} is reduced to the simple gap equation \cite{bcs}
\begin{equation}\label{eqn:delta1}
1=U_1\int_0^{\hslash\omega_D}
 \frac{1}{\,\sqrt{\,\xi^2+\Delta_1(T)^2\,}\,}\,
 \tanh \frac{\, \sqrt{\,\xi^2+\Delta_1(T)^2\,}\,}{2T}\,d\xi, \quad 0 \leq T \leq\tau_1 \, .
\end{equation}
Here the temperature $\tau_1>0$ is defined by (see \cite{bcs})
\[
1=U_1\int_0^{\hslash\omega_D}
\frac{1}{\,\xi\,}\,\tanh \frac{\xi}{\,2\tau_1\,}\,d\xi.
\]
See also Niwa \cite{niwa} and Ziman \cite{ziman}. The implicit function theorem implies the following.
\begin{proposition}[{\cite[Proposition 2.2]{watanabe1}}]
Set
\[
\Delta=\frac{\, \hslash\omega_D \,}{\,\sinh\frac{1}{\,U_1\,}\,}.
\]
Then there is a unique nonnegative solution $\Delta_1: [\,0,\,\tau_1\,] \to [0,\,\infty)$ to the simple gap equation such that the solution $\Delta_1$ is continuous and strictly decreasing with respect to the temperature on the closed interval $[\,0,\,\tau_1\,]$:
\[
\Delta_1(0)=\Delta>\Delta_1(T_1)>\Delta_1(T_2)>\Delta_1(\tau_1)=0, \qquad 0<T_1<T_2<\tau_1.
\]
Moreover, the solution $\Delta_1$ is of class $C^2$ on the interval $[\,0,\,\tau_1\,)$ and satisfies
\[
\Delta_1'(0)=\Delta_1''(0)=0 \quad \mbox{and} \quad \lim_{T\uparrow \tau_1} \Delta_1'(T)=-\infty.
\]
\end{proposition}

\begin{remark}
We set $\Delta_1(T)=0$ for $T>\tau_1$.
\end{remark}

We introduce another positive constant $U_2>0$. Let $0<U_1<U_2$ and set $U(x,\,\xi)=U_2$ at all $(x,\,\xi) \in [0,\, \hslash\omega_D]^2$. Then an argument similar to that in the proposition just above implies that there is a unique nonnegative solution $\Delta_2: [\,0,\,\tau_2\,] \to [0,\,\infty)$ to the simple gap equation
\begin{equation}\label{eqn:delta2}
1=U_2\int_0^{\hslash\omega_D}
 \frac{1}{\,\sqrt{\,\xi^2+\Delta_2(T)^2\,}\,}\,
 \tanh \frac{\, \sqrt{\,\xi^2+\Delta_2(T)^2\,}\,}{2T}\,d\xi, \qquad
0\leq T\leq \tau_2.
\end{equation}
Here, $\tau_2>0$ is defined by
\[
1=U_2\int_0^{\hslash\omega_D}
\frac{1}{\,\xi\,}\,\tanh \frac{\xi}{\,2\tau_2\,}\,d\xi.
\]

We again set $\Delta_2(T)=0$ for $T>\tau_2$. A straightforward calculation gives the following.

\begin{lemma}[{\cite[Lemma 1.5]{watanabe2}}] \quad {\rm (a)} The inequality $\tau_1<\tau_2$ holds.

\noindent {\rm (b)} If \   $0\leq T<\tau_2$, then $\Delta_1(T)<\Delta_2(T)$. If \  $T\geq \tau_2$, then $\Delta_1(T)=\Delta_2(T)=0$.
\end{lemma}

See figure 1.  The function $\Delta_2$ has properties similar to those of the function $\Delta_1$.

\begin{figure}[htbp]\hspace{3cm}
\includegraphics[width=8cm]{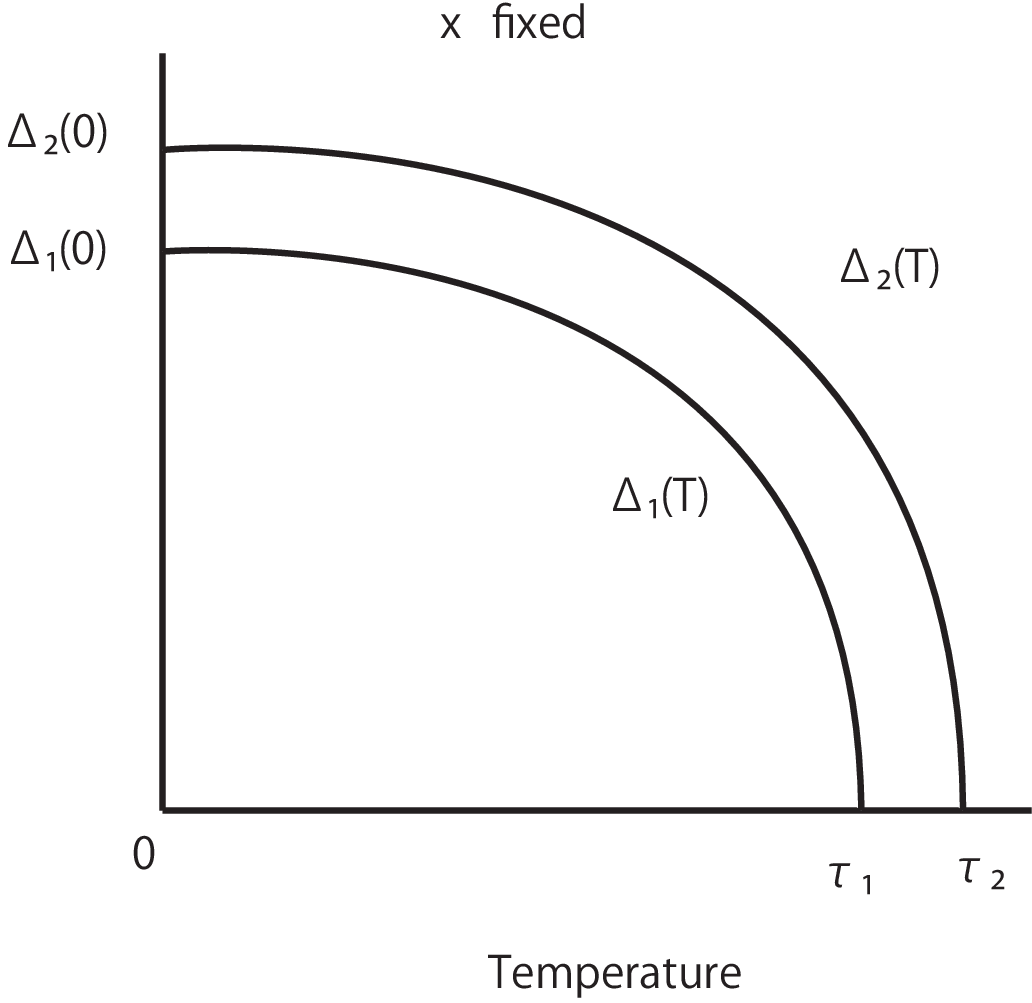}
\caption{\textsf{The graphs of the functions $\Delta_1$ and $\Delta_2$ with $x$ fixed.}}
\end{figure}

We then deal with the BCS gap equation \eqref{eqn:bcsgapeq}, where the potential $U$ is not a constant but a function. We assume the following condition on $U$:
\begin{equation}\label{eqn:conditionU}
U_1 \leq U(x,\,\xi) \leq U_2 \quad \mbox{at all} \quad (x,\,\xi) \in [0,\, \hslash\omega_D]^2, \qquad U(\cdot,\,\cdot) \in C([0,\, \hslash\omega_D]^2).
\end{equation}
Let $0 \leq T \leq \tau_2$ and fix $T$. We now consider the Banach space $C[0,\, \hslash\omega_D]$ consisting of continuous functions of $x$ only, and deal with the following temperature dependent subset $V_T$:
\[
V_T=\left\{ u(T,\,\cdot) \in C[0,\, \hslash\omega_D]: \; \Delta_1(T) \leq u(T,\,x) \leq \Delta_2(T) \; \mbox{at} \; x \in [0,\, \hslash\omega_D] \right\}.
\]

Applying the Schauder fixed-point theorem to our operator \eqref{eqn:ouroperator} defined on $V_T$ implies the following.
\begin{theorem}[{\cite[Theorem 2.2]{watanabe2}}] \label{thm:3-1}
Assume \eqref{eqn:conditionU} and fix $T \in [0,\, \tau_2]$. Then there is a unique nonnegative solution $u_0(T,\,\cdot) \in V_T$ to the BCS gap equation \eqref{eqn:bcsgapeq}:
\[
u_0(T,\, x)=\int_0^{\hslash\omega_D}
\frac{U(x,\,\xi)\, u_0(T,\, \xi)}{\,\sqrt{\,\xi^2+u_0(T,\, \xi)^2\,}\,}\,
\tanh \frac{\,\sqrt{\,\xi^2+u_0(T,\, \xi)^2\,}\,}{2T}\, d\xi, \, \quad  x \in [0,\, \hslash\omega_D].
\]
Consequently, the solution is continuous with respect to $x$ and varies with the temperature as follows:
\[
\Delta_1(T) \leq u_0(T,\, x) \leq \Delta_2(T) \quad \mbox{at} \quad
(T,\,x) \in [0,\, \tau_2] \times [0,\, \hslash\omega_D].
\]
\end{theorem}

See figure 2.

\begin{figure}[htbp]
\includegraphics[width=12cm]{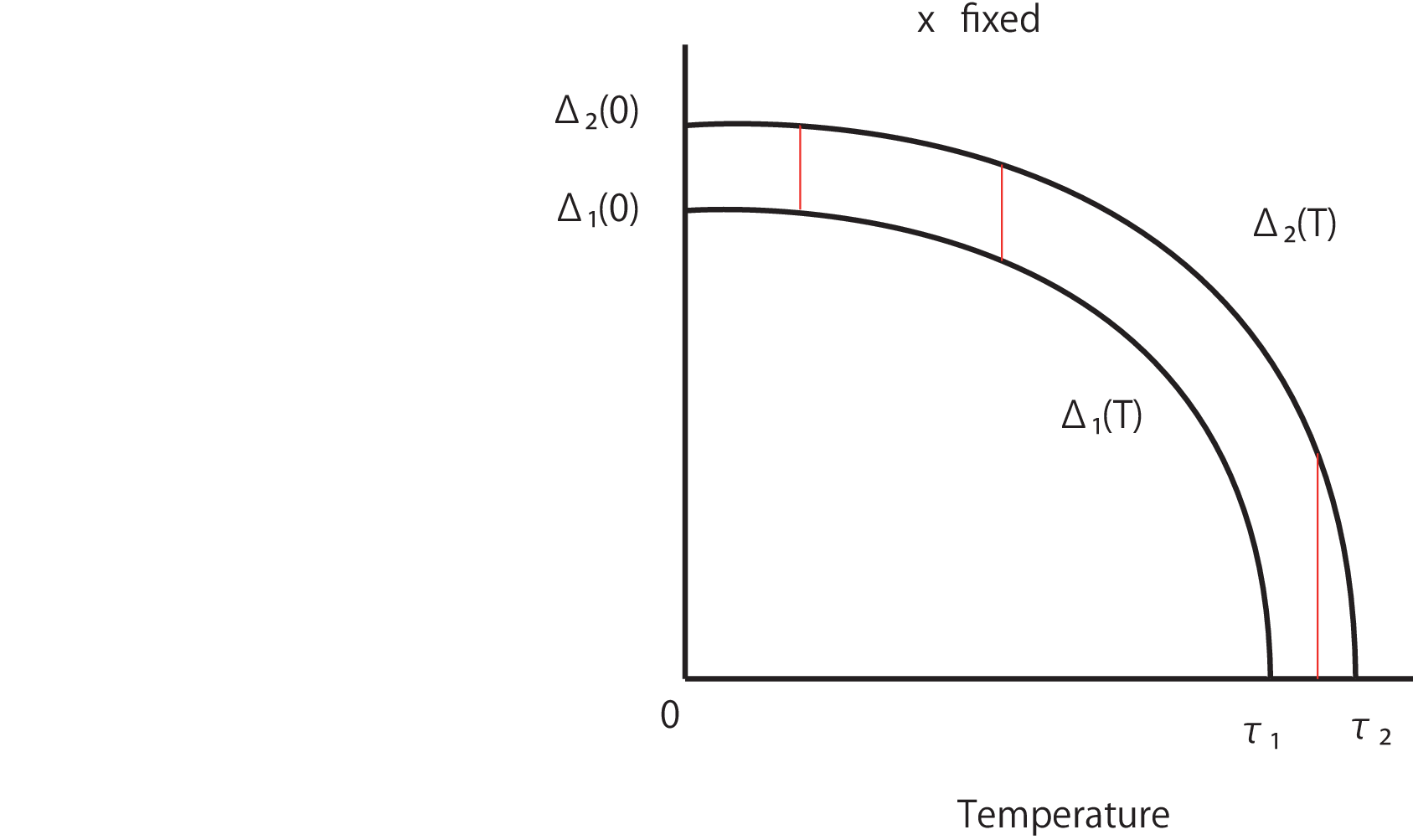}
\caption{\textsf{For each $T$, the solution $u_0(T,\, x)$ lies between $\Delta_1(T)$ and $\Delta_2(T)$.}}
\end{figure}

Let $U_0 >0$ be a positive constant satisfying $U_0 < U_1 < U_2$. An argument similar to that in the proposition above implies that there is a unique nonnegative solution $\Delta_0: [\,0,\,\tau_0'\,] \to [0,\,\infty)$ to the simple gap equation
\[
1=U_0\int_0^{\hslash\omega_D}
 \frac{1}{\,\sqrt{\,\xi^2+\Delta_0(T)^2\,}\,}\,
 \tanh \frac{\, \sqrt{\,\xi^2+\Delta_0(T)^2\,}\,}{2T}\,d\xi, \qquad
0\leq T\leq \tau_0'.
\]
Here, $\tau_0'>0$ is defined by
\[
1=U_0\int_0^{\hslash\omega_D}
\frac{1}{\,\xi\,}\,\tanh \frac{\xi}{\,2\tau_0'\,}\,d\xi.
\]
We set $\Delta_0(T)=0$ for $T>\tau_0'$.

\begin{remark}
Let the functions $\Delta_l$ $(l=1,\, 2)$ and $\Delta_0$ be as above. For each function, there is its inverse. We have $\Delta_l^{-1}: \, [0,\,\Delta_l(0)] \to [0,\, \tau_l]$ and $\Delta_0^{-1}: \, [0,\,\Delta_0(0)] \to [0,\, \tau_0']$. Here,
\[
\Delta_l(0)=\frac{\, \hslash\omega_D \,}{\,\sinh\frac{1}{\,U_l\,}\,}, \quad \Delta_0(0)=\frac{\, \hslash\omega_D \,}{\,\sinh\frac{1}{\,U_0\,}\,}
\]
and $\Delta_0(0)<\Delta_1(0)<\Delta_2(0)$. See \cite{watanabe3} for more details.
\end{remark}

We introduce another temperature $T_1$. Let $T_1$ satisfy \   $\displaystyle{ 
0<T_1<\Delta_0^{-1}\left( \frac{\, \Delta_0(0)\,}{2} \right) }$ and
\[
\frac{\,\Delta_0(0)\,}{\, 4\, \Delta_2^{-1}\left( \Delta_0(T_1) \right) \,} \tanh \frac{\,\Delta_0(0)\,}{\, 4\, \Delta_2^{-1}\left( \Delta_0(T_1) \right) \,} > \frac{1}{\, 2\,}\left( 1+\frac{\, 4\hslash^2\omega_D^2\,}{\,\Delta_0(0)^2\,} \right).
\]

\begin{remark}
The value of the temperature $T_1$ is very small by experimental values in many superconductors. See figure 3.
\end{remark}

We consider the following subset $V$ of the Banach space $C([0,\, T_1] \times [0,\,\hslash\omega_D])$ consisting of continuous functions both of the temperature $T$ and of the energy $x$:
\begin{eqnarray}
V &=& \left\{ u \in C([0,\, T_1] \times [0,\,\hslash\omega_D]): \Delta_1(T) \leq u(T,\,x) \leq \Delta_2(T) \right.  \nonumber \\
 & & \qquad \qquad \left. at \; \; (T,\,x) \in [0,\, T_1] \times [0,\,\hslash\omega_D] \right\}. \nonumber
\end{eqnarray}

Applying the Banach fixed-point theorem to our operator \eqref{eqn:ouroperator} defined on $V$ implies the following.
\begin{theorem}[{\cite[Theorem 1.2]{watanabe3}}]
Assume \eqref{eqn:conditionU}. Let $u_0$, $T_1$ and $V$ be as above. Then $u_0 \in V$. Consequently, the solution  $u_0$ to the BCS gap equation \eqref{eqn:bcsgapeq} is continuous on $[0,\, T_1] \times [0,\,\hslash\omega_D]$.
\end{theorem}

See figure 3.

\begin{figure}[htbp]
\includegraphics[width=12cm]{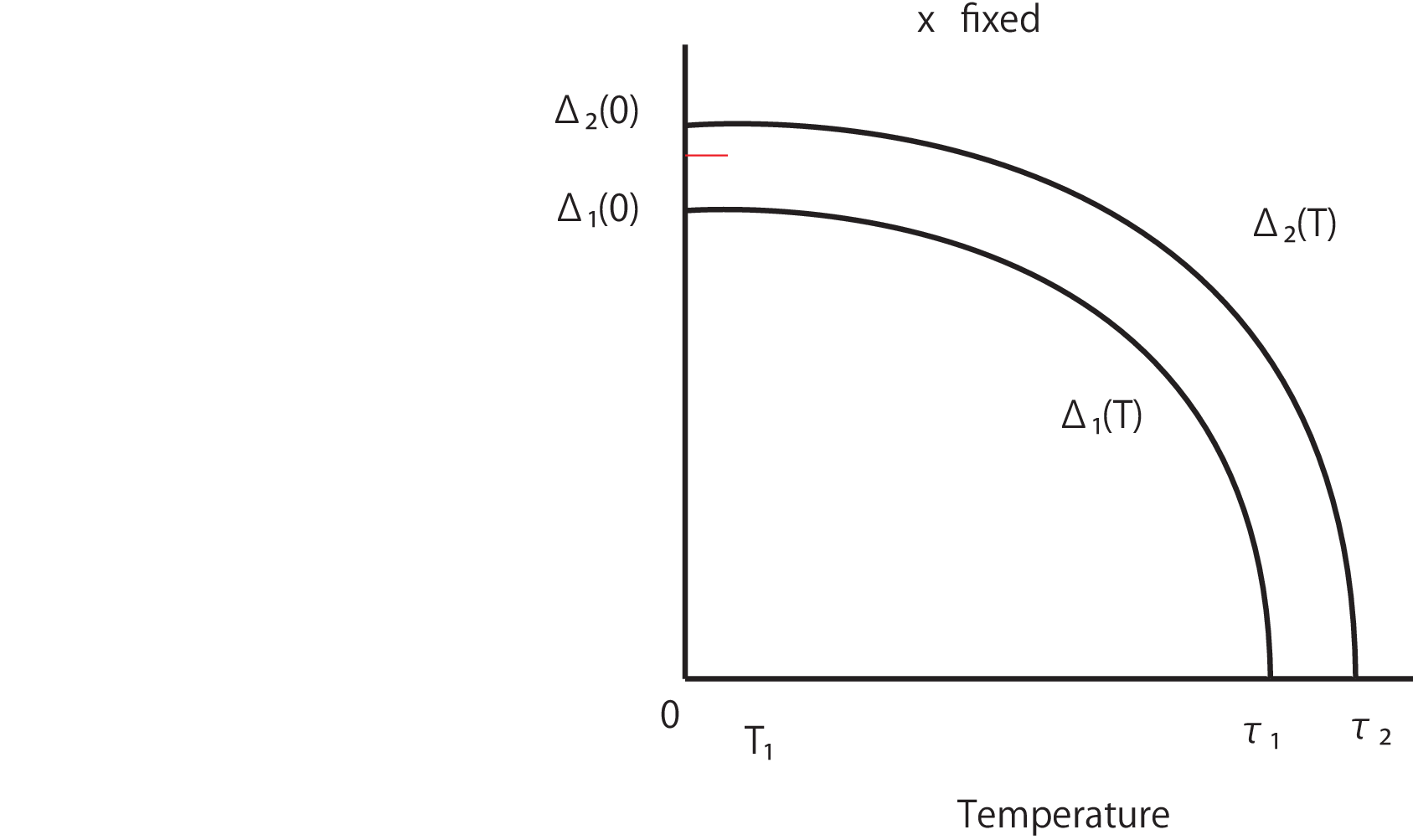}
\caption{\textsf{The solution $u_0$ is continuous on $[0,\, T_1] \times [0,\,\hslash\omega_D]$.}}
\end{figure}

Let us denote by $z_0 >0$ a unique solution to the equation $\displaystyle{ \frac{2}{\,  z \,} = \tanh z }$
\quad  $(z>0)$. Note that $z_0$ is nearly equal to 2.07 and that $\displaystyle{ \frac{2}{\,  z \,} \leq \tanh z }$ for $z \geq z_0$ . Let  $\tau_0 \, (>0)$ be a temperature satisfying
\begin{equation}\label{eq:tau0}
\Delta_1(\tau_0) = 2z_0\tau_0 \,.
\end{equation}
From \eqref{eq:tau0} it follows immediately that $(0 <)\, \tau_0 <\tau_1$ .

\begin{remark}
As compared with $T_1$ in the preceding section, the temperature $\tau_0$ is not very small and is nearly equal to $T_c/2$ by experimental values in many superconductors. See figure 4. Here, $T_c$ is the transition temperature satisfying $\tau_1\leq T_c \leq \tau_2$ , and divides normal conductivity and superconductivity.
\end{remark}

Let $0<\tau < \tau_0$ and fix $\tau$. We then deal with the following subset $W$ of the Banach space $C([0,\, \tau] \times [0,\,\hslash\omega_D])$:
\begin{eqnarray}
W &=& \left\{ u \in C([0,\, \tau] \times [0,\,\hslash\omega_D]) :  0 \leq u(T,\,x)-u(T',\,x) \leq \gamma \left( T'-T \right) \; \; (T<T'),
\right.  \nonumber \\
& & \left. \Delta_1(T) \leq u(T,\,x) \leq \Delta_2(T) \; \; at \; \; (T,\,x), \; (T',\,x) \in [0,\, \tau] \times [0,\,\hslash\omega_D] \right\}. \nonumber
\end{eqnarray}
Here, $\gamma$ is defined by \eqref{eqn:gamma} below. The following is our main result.

\begin{theorem}\label{thm:main}
Assume \eqref{eqn:conditionU}. Let $\tau$ and $W$ be as above. Then the operator $A:\,  W \to W$ has a unique fixed point $u_0 \in W$, i.e., there is a unique nonnegative solution $u_0 \in W$ to the BCS gap equation \eqref{eqn:bcsgapeq}. Consequently, the solution $u_0$ is continuous on $[0,\, \tau] \times [0,\,\hslash\omega_D]$, and is Lipschitz continuous and monotonically decreasing with respect to the temperature $T$. Moreover, it satifies $\Delta_1(T) \leq u_0(T,\,x) \leq \Delta_2(T)$ at $(T,\,x) \in [0,\, \tau] \times [0,\,\hslash\omega_D]$.
\end{theorem}

See figure 4.

\begin{figure}[htbp]\hspace{3cm}
\includegraphics[width=8cm]{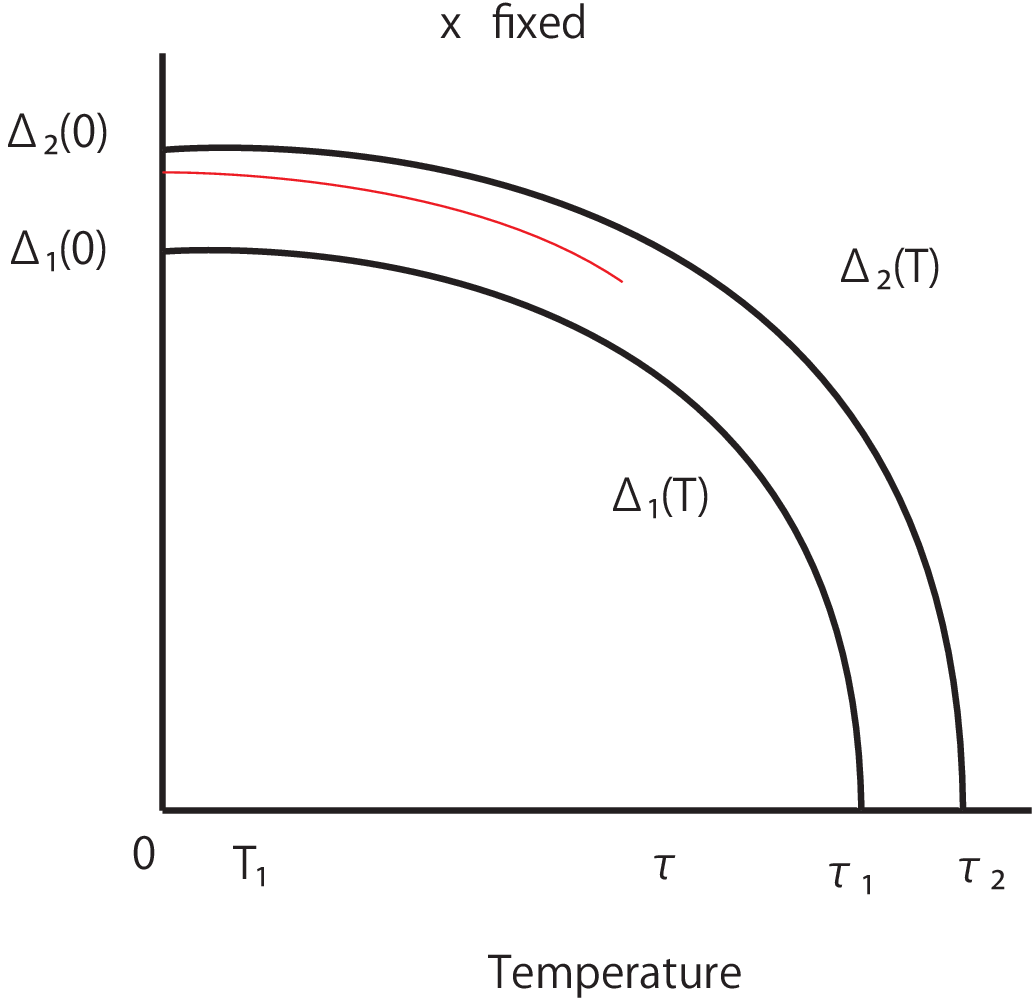}
\caption{\textsf{The solution $u_0$ belongs to the subset $W$.}}
\end{figure}

\section{Proof of Theorem \ref{thm:main}}

We prove Theorem \ref{thm:main} in a sequence of lemmas.

\begin{lemma}\label{lm:functionF}
Let $0<\tau < \tau_0$ and fix $\tau$. Define a function $F$ on $[0, \, \tau]$ by
\[
F(T)=\int_0^{\hslash\omega_D} \frac{1}{\,\sqrt{\,\xi^2+\Delta_1(T)^2\,}\,}\,
 \tanh \frac{\, \sqrt{\,\xi^2+\Delta_1(T)^2\,}\,}{2\tau_0}\,d\xi, \qquad T \in [0,\, \tau].
\]
Then the function $F$ is continuous on $[0, \, \tau]$.
\end{lemma}

\begin{proof}
Let $T \in [0,\, \tau]$. Note that $\displaystyle{ \frac{z}{\, \cosh^2z \,} \leq \tanh z }$ \  $(z \geq 0)$ and that $\displaystyle{ \frac{\,\tanh z \,}{z} \leq 1 }$ \  $(z \geq 0)$. Then
\begin{eqnarray*}
\, & & \left| F(T+h)-F(T) \right| \nonumber \\
&\leq& \int_0^{\hslash\omega_D}
 \frac{\, \left| \Delta_1(T+h)^2-\Delta_1(T)^2 \right| \,}{\,2\left( \,\xi^2+d \,\right)^{3/2}\,}\, \left\{
 \tanh \frac{\, \sqrt{\,\xi^2+d \,}\,}{2\tau_0}+\frac{\, \sqrt{\, \xi^2+d\,}\,}{2\tau_0}
\frac{1}{\, \cosh^2 \frac{\, \sqrt{\,\xi^2+d \,}\,}{2\tau_0} \,} \right\}
\,d\xi  \nonumber \\
&\leq& \left| \Delta_1(T+h)^2-\Delta_1(T)^2 \right| \, \int_0^{\hslash\omega_D}
 \frac{1}{\,\left( \,\xi^2+d \,\right)^{3/2}\,}\, 
 \tanh \frac{\, \sqrt{\,\xi^2+d \,}\,}{2\tau_0}\,d\xi  \nonumber \\
&\leq& \left| \Delta_1(T+h)^2-\Delta_1(T)^2 \right| \, \int_0^{\hslash\omega_D}
 \frac{d\xi}{\,2\tau_0\left( \,\xi^2+d \,\right)\,} \, . \nonumber \\
\end{eqnarray*}
Here, $d$ satisfies $\Delta_1(T+h)^2<d<\Delta_1(T)^2$ or $\Delta_1(T+h)^2>d>\Delta_1(T)^2$.
Since $d\geq \Delta_1(\tau)^2$, it follows that
\[
\left| F(T+h)-F(T) \right| \leq \left| \Delta_1(T+h)^2-\Delta_1(T)^2 \right|
 \frac{1}{\, 2\tau_0\Delta_1(\tau) \,} \, \arctan \frac{\, \hslash\omega_D \,}{\Delta_1(\tau)}.
\]
Continuity of the function $\Delta_1$ proves the lemma.
\end{proof}

Let $0<\tau < \tau_0$ and fix $\tau$. In view of Lemma \ref{lm:functionF}, we set
\begin{eqnarray}
a&=&\max_{0 \leq T \leq \tau} \int_0^{\hslash\omega_D}
 \frac{1}{\,\sqrt{\,\xi^2+\Delta_1(T)^2\,}\,}\,
 \tanh \frac{\, \sqrt{\,\xi^2+\Delta_1(T)^2\,}\,}{2\tau_0}\,d\xi, \label{eq:a} \\
b&=&\frac{32\tau^2}{\, \Delta_1(\tau)^2 \,} \, \arctan \frac{\, \hslash\omega_D \,}{\Delta_1(\tau)}.
\nonumber
\end{eqnarray}
Then, for $T \in [0,\, \tau]$, 
\begin{eqnarray*}
1&=&U_1\int_0^{\hslash\omega_D}
 \frac{1}{\,\sqrt{\,\xi^2+\Delta_1(T)^2\,}\,}\,
 \tanh \frac{\, \sqrt{\,\xi^2+\Delta_1(T)^2\,}\,}{2T}\,d\xi \nonumber \\
&>&U_1\int_0^{\hslash\omega_D}
 \frac{1}{\,\sqrt{\,\xi^2+\Delta_1(T)^2\,}\,}\,
 \tanh \frac{\, \sqrt{\,\xi^2+\Delta_1(T)^2\,}\,}{2\tau_0}\,d\xi \nonumber \\
\end{eqnarray*}
by \eqref{eqn:delta1}. Lemma \ref{lm:functionF} implies $1>U_1a$, where $a$ is that in
\eqref{eq:a}. We choose $U_2 \, (>U_1)$ such that $1>U_2a$ holds true. Set
\begin{equation}\label{eqn:gamma}
\gamma = \frac{\, U_2b\,}{\, 1-U_2a \,} \quad (>0).
\end{equation}

A straightforward calculation gives the following.
\begin{lemma}\label{lm:setw}
The subset $W$ is bounded, closed, convex and nonempty.
\end{lemma}

\begin{lemma}\label{lm:delta}
If $u \in W$, then $\Delta_1(T) \leq Au(T,\,x) \leq \Delta_2(T)$ at all $(T,\,x) \in [0,\, \tau] \times [0,\,\hslash\omega_D]$.
\end{lemma}

\begin{proof}
Since $u(T,\,x) \leq \Delta_2(T)$, it follows that
\[
\frac{ u(T,\, \xi) }{ \,\sqrt{\,\xi^2+u(T,\, \xi)^2\, }\, } \leq
\frac{ \Delta_2(T) }{ \,\sqrt{\,\xi^2+ \Delta_2(T)^2\, }\, }.
\]
Therefore \eqref{eqn:delta2} gives
\[
Au(T,\,x) \leq U_2\int_0^{\hslash\omega_D} \frac{ \Delta_2(T) }{\,\sqrt{\,\xi^2+\Delta_2(T)^2\,}\,}\,
 \tanh \frac{\, \sqrt{\,\xi^2+\Delta_2(T)^2\,}\,}{2T}\,d\xi = \Delta_2(T).
\]
Similarly we can show the rest.
\end{proof}

\begin{lemma}\label{lm:G}
Let $T \in [0,\,\tau_0]$ and let $X \in [\Delta_1(\tau_0)^2,\,\infty)$. Define a function $G$ by
\[
G(T,\, X,\,\xi)=\xi^2\tanh \frac{\, \sqrt{\,\xi^2+X \,}\,}{2T}+\frac{4XT}{\, \sqrt{\,\xi^2+X \,}\,}, \quad
0 \leq \xi \leq \hslash\omega_D\, .
\]
Then $G$ is a monotone increasing function with respect to $T \in [0,\,\tau_0]$. Consequently, $G(T,\, X,\,\xi) \leq G(\tau_0,\, X,\,\xi)$.
\end{lemma}

\begin{proof}
A straightforward calculation gives
\[
\frac{\, \partial G \,}{\partial T}=\frac{1}{\,  2\sqrt{\xi^2+X}\,}
\left(  \sqrt{8X}+\frac{\xi\sqrt{\xi^2+X}}{\,  T\cosh \frac{\sqrt{\xi^2+X}}{2T}\,}  \right)
\left(  \sqrt{8X}-\frac{\xi\sqrt{\xi^2+X}}{\,  T\cosh \frac{\sqrt{\xi^2+X}}{2T}\,}  \right).
\]
Since $\displaystyle{ \frac{z}{\, \cosh z \,} \leq \frac{2}{\, z\,} }$ \  $(z \geq 0)$, it follows from
\eqref{eq:tau0} that
\begin{eqnarray*}
\sqrt{8X}-\frac{\xi\sqrt{\xi^2+X}}{\,  T\cosh \frac{\sqrt{\xi^2+X}}{2T}\,} &\geq&
\sqrt{8X}-\frac{8\xi T}{\,  \sqrt{\xi^2+X} \,} \nonumber \\
&=& \frac{\sqrt{8X}\sqrt{\xi^2+X}-8\xi T}{\,  \sqrt{\xi^2+X} \,} \nonumber \\
&\geq& \frac{\sqrt{8}\, \Delta_1(\tau_0)\xi-8\xi \tau_0}{\,  \sqrt{\xi^2+X} \,} \nonumber \\
&=& \frac{2\sqrt{8}\xi \tau_0}{\,  \sqrt{\xi^2+X} \,}\left( z_0-\sqrt{2} \right) \nonumber \\
&\geq& 0.
\end{eqnarray*}
Note that $\sqrt{X} \geq \Delta_1(\tau_0)$ and that $z_0$ is nearly equal to 2.07. The result thus follows.
\end{proof}

\begin{lemma}\label{lm:gamma}
Let $0<\tau < \tau_0$ and fix $\tau$. For $T, \, T' \in [0,\, \tau]$, let $T<T'$. If $u \in W$, then
\[
0 \leq Au(T,\,x)-Au(T',\,x) \leq \gamma \left( T'-T \right), \qquad x \in [0,\,\hslash\omega_D].
\]
\end{lemma}

\begin{proof} \   Step 1. We first show $Au(T,\,x)-Au(T',\,x) \geq 0$. 
\[
Au(T,\,x)-Au(T',\,x)=\int_0^{\hslash\omega_D} U(x,\,\xi) \left( K_1 + K_2 \right) \, d\xi,
\]
where
\begin{eqnarray}
K_1 &=& \frac{u(T,\, \xi)}{\,\sqrt{\,\xi^2+u(T,\, \xi)^2\,}\,}\tanh \frac{\,\sqrt{\,\xi^2+u(T,\, \xi)^2\,}\,}{2T}
-\frac{u(T',\, \xi)}{\,\sqrt{\,\xi^2+u(T',\, \xi)^2\,}\,}\tanh \frac{\,\sqrt{\,\xi^2+u(T',\, \xi)^2\,}\,}{2T} \,,
\nonumber \\
K_2 &=& \frac{u(T',\, \xi)}{\,\sqrt{\,\xi^2+u(T',\, \xi)^2\,}\,}\left\{
\tanh \frac{\,\sqrt{\,\xi^2+u(T',\, \xi)^2\,}\,}{2T}
-\tanh \frac{\,\sqrt{\,\xi^2+u(T',\, \xi)^2\,}\,}{2T'} \right\}\,.  \nonumber
\end{eqnarray}
Since $u(T,\,\xi) \geq u(T',\,\xi)$, it follows that
\[
\frac{u(T,\, \xi)}{\,\sqrt{\,\xi^2+u(T,\, \xi)^2\,}\,} \geq \frac{u(T',\, \xi)}{\,\sqrt{\,\xi^2+u(T',\, \xi)^2\,}\,}.
\]
Hence $K_1\geq 0$. Clearly, $K_2\geq 0$. Thus $Au(T,\,x)-Au(T',\,x)\geq 0$.

Step 2. \   We next show $Au(T,\,x)-Au(T',\,x) \leq \gamma \left( T'-T \right)$. \   Since $\displaystyle{ \frac{z}{\,\cosh^2z\,} \leq \frac{2}{\,  z\,}}$ \   $(z \geq 0)$, it follows from Lemma \ref{lm:G} that
\begin{eqnarray*}
K_1&=&\frac{1}{\,\left( \xi^2+c^2 \right)^{3/2} \,}\left\{ \xi^2\tanh
\frac{\,\sqrt{\,\xi^2+c^2\,}\,}{2T} + \frac{\, c^2\, \sqrt{\,\xi^2+c^2\,}\,}{\, 2T\,\cosh^2 \frac{\,\sqrt{\,\xi^2+c^2\,}\,}{2T}\,} \right\} \left\{ u(T,\,\xi)-u(T',\,\xi) \right\} \nonumber \\
&\leq& \frac{1}{\,\left( \xi^2+c^2 \right)^{3/2} \,} \, G(T,\, c^2,\, \xi) \, \gamma (T'-T) \nonumber \\
&\leq& \frac{1}{\,\left( \xi^2+c^2 \right)^{3/2} \,} \, G(\tau_0,\, c^2,\, \xi) 
\, \gamma (T'-T), \nonumber \\
\end{eqnarray*}
where $c$ depends on $T$, $T'$, $\xi$ and $u$, and satisfies $u(T,\,\xi) >c>u(T',\,\xi)$. Note that
\[
\frac{\,\sqrt{\,\xi^2+c^2\,}\,}{2\tau_0} \geq \frac{\,\sqrt{\, c^2\,}\,}{2\tau_0} >
\frac{\,\Delta_1(\tau_0)\,}{2\tau_0}=z_0 \, .
\]
The substitution $\displaystyle{ z=\frac{\,\sqrt{\,\xi^2+c^2\,}\,}{2\tau_0} }$ therefore turns \   $\displaystyle{ \frac{2}{\,  z \,} \leq \tanh z }$ \    $(z \geq z_0)$ \   into
\[
\frac{4\tau_0}{\,\sqrt{\,\xi^2+c^2\,}\,} \leq \tanh \frac{\,\sqrt{\,\xi^2+c^2\,}\,}{2\tau_0}.
\]
Hence
\begin{eqnarray*}
K_1 &\leq& \frac{1}{\,\sqrt{\,\xi^2+c^2\,}\,} \tanh \frac{\,\sqrt{\,\xi^2+c^2\,}\,}{2\tau_0}
\, \gamma (T'-T) \nonumber \\
&\leq& \frac{1}{\,\sqrt{\,\xi^2+\Delta_1(T')^2\,}\,} \tanh \frac{\,\sqrt{\,\xi^2+\Delta_1(T')^2\,}\,}{2\tau_0} \, \gamma (T'-T). \nonumber
\end{eqnarray*}
Since 
$\displaystyle{ \frac{z}{\,\cosh z\,} \leq \frac{\,2\,}{z}}$ \   $(z \geq 0)$, it follows that
\begin{eqnarray*}
K_2 &=& \frac{ \, 2u(T',\,\xi)(T'-T) \,}{ \xi^2+u(T',\,\xi)^2 } \left\{
\frac{\,\sqrt{\,\xi^2+u(T',\,\xi)^2\,}\,}{2T''} \frac{1}{\, \cosh \frac{\,\sqrt{\,\xi^2+u(T',\,\xi)^2\,}\,}{2T''}  \, } 
 \right\}^2  \nonumber \\
&\leq& \frac{ \, 2u(T',\,\xi)(T'-T) \,}{ \xi^2+u(T',\,\xi)^2 }\frac{16(T'')^2}{\, \xi^2+u(T',\,\xi)^2} 
 \nonumber \\
&\leq& \frac{ \,  (T'-T)\,  32\tau^2\,  }{ \Delta_1(\tau)  }\frac{1}{\, \xi^2+\Delta_1(\tau)^2},
\end{eqnarray*}
where $T<T''<T'$. Thus
\begin{eqnarray*}
& & Au(T,\,x)-Au(T',\,x) \nonumber \\
&\leq& (T'-T) \, U_2\int_0^{\hslash\omega_D} \left( 
\frac{\gamma}{\,\sqrt{\,\xi^2+\Delta_1(T')^2\,}\,}
\tanh \frac{\,\sqrt{\,\xi^2+\Delta_1(T')^2\,}\,}{2\tau_0} 
+\frac{ \,  32\tau^2\,  }{ \, \Delta_1(\tau) \,  }\frac{1}{\, \xi^2+\Delta_1(\tau)^2}
\right) \, d\xi \nonumber \\
&\leq&  (T'-T) \, U_2 \left( \gamma \, a+b  \right) \nonumber \\
&=&  \gamma \, (T'-T). \nonumber \\
\end{eqnarray*}
\end{proof}

\begin{lemma}\label{lm:contin}
If $u \in W$, then $Au \in C([0,\, \tau] \times [0,\,\hslash\omega_D])$.
\end{lemma}

\begin{proof}
Let $T<T'$. Then
\begin{equation}\label{eqn:au}
\left| Au(T,\,x)-Au(T',\,x') \right| \leq \left| Au(T,\,x)-Au(T',\,x) \right| +
\left| Au(T',\,x)-Au(T',\,x') \right|.
\end{equation}
Since $U(\cdot, \,\cdot)$ is uniformly continuous, for an arbitrary $\varepsilon>0$, there is a $\delta_1>0$ such that
$|x-x'|<\delta_1$ implies
\[
\left| U(x,\, \xi)-U(x',\, \xi) \right| <\frac{\varepsilon}{\, 2\hslash\omega_D \,}.
\]
Note that the $\delta_1>0$ depends neither on $x$, nor on $x'$, nor on $\xi$, nor on $u \in W$. Hence the second term on the right of \eqref{eqn:au} becomes
\[
\left| Au(T',\,x)-Au(T',\,x') \right| \leq \int_0^{\hslash\omega_D} \left| U(x,\, \xi)-U(x',\, \xi) \right| \,d\xi
<\frac{\,\varepsilon\,}{2}.
\]
On the other hand, the first term on the right of \eqref{eqn:au} becomes
\[
\left| Au(T,\,x)-Au(T',\,x) \right| \leq \gamma (T'-T)<\frac{\,\varepsilon\,}{2}
\]
by the preceding lemma. Here, $T'-T<\varepsilon/(2\gamma)$. Thus
\[
\left| Au(T,\,x)-Au(T',\,x') \right|<\varepsilon, \quad
 (T'-T)+\left| x-x' \right|<\delta=\min\left( \delta_1,\, \frac{\varepsilon}{\,  2\gamma \,} \right).
\]
Note that the $\delta>0$ depends neither on $x$, nor on $x'$, nor on $\xi$, nor on $u \in W$, nor on $T$, nor on $T'$.
\end{proof}

The preceding three lemmas imply the following.
\begin{lemma} \quad 
$\displaystyle{ AW \subset W}$.
\end{lemma}

\begin{lemma}\label{lm:setaw} \quad 
The set $AW$ is relatively compact.
\end{lemma}

\begin{proof}
Let $u \in W$. Lemma \ref{lm:delta} then implies 
\[
Au(T,\, x)\leq \Delta_2(0)=\frac{\, \hslash\omega_D \,}{\,\sinh\frac{1}{\,U_2\,}\,}.
\]
So the set $AW$ is uniformly bounded. As mentioned in the proof of Lemma \ref{lm:contin}, the $\delta$ does not depend on $u \in W$. Hence the set $AW$ is equicontinuous. The result thus follows from the Ascoli--Arzel$\grave{\mbox{a}}$ theorem.
\end{proof}

\begin{lemma}\label{lm:acontinuous}
The operator $A:\,  W \to W$ is continuous.
\end{lemma}

\begin{proof}
Let $u,\, v \in W$. Then combining an argument similar to that in the proof of Lemma \ref{lm:gamma} with \eqref{eqn:delta1} gives
\begin{eqnarray}
& & \left| Au(T,\,x)-Av(T,\,x) \right| \nonumber \\
&\leq& U_2 \int_0^{\hslash\omega_D} \frac{1}{\,\left( \xi^2+d^2 \right)^{3/2} \,}
\left\{ \xi^2\tanh \frac{\,\sqrt{\,\xi^2+d^2\,}\,}{2T} + \frac{\, d^2\, \sqrt{\,\xi^2+d^2\,}\,}{\, 2T\,\cosh^2 \frac{\,\sqrt{\,\xi^2+d^2\,}\,}{2T}\,} \right\} \left| u(T,\,\xi)-v(T,\,\xi) \right| \, d\xi \nonumber \\
&\leq& U_2 \int_0^{\hslash\omega_D}  \frac{1}{\,\sqrt{ \xi^2+d^2 } \,} 
\tanh \frac{\,\sqrt{\,\xi^2+d^2\,}\,}{2T} \, d\xi \, \left\| u-v \right\| \nonumber \\
&\leq& \frac{\, U_2\,}{U_1} \int_0^{\hslash\omega_D}  \frac{U_1}{\,\sqrt{ \xi^2+\Delta_1(T)^2 } \,} 
\tanh \frac{\,\sqrt{\,\xi^2+\Delta_1(T)^2\,}\,}{2T} \, d\xi \, \left\| u-v \right\| \nonumber \\
&=& \frac{\, U_2\,}{U_1} \, \left\| u-v \right\|. \nonumber
\end{eqnarray}
Here, $d$ satisfies $u(T,\,\xi)<d<v(T,\,\xi)$ or $u(T,\,\xi)>d>v(T,\,\xi)$, and $\| \cdot \|$ denotes the norm of the Banach space $C([0,\, \tau] \times [0,\,\hslash\omega_D])$. The result thus follows.
\end{proof}

Lemmas \ref{lm:setw}, \ref{lm:setaw} and \ref{lm:acontinuous} imply the following.

\begin{lemma} \quad 
The operator $A:\,  W \to W$ is compact.
\end{lemma}

The uniqueness of the nonzero fixed point of $A:\,  W \to W$ was pointed out in Theorem \ref{thm:3-1}. The Schauder fixed-point theorem thus proves Theorem \ref{thm:main}.

\section*{Acknowledgments}
S. Watanabe is supported in part by the JSPS Grant-in-Aid for Scientific Research (C) 24540112.

\end{document}